# Deep-learning-enabled Brain Hemodynamic Mapping Using Resting-state fMRI

Xirui Hou[1,2*], Pengfei Guo[3], Puyang Wang[4], Peiying Liu[5], Doris D.M. Lin[2], Hongli Fan[1,2], Yang Li[2], Zhiliang Wei[2,6], Zixuan Lin[2], Dengrong Jiang[2], Jin Jin[7], Catherine Kelly[8], Jay J. Pillai[2,9], Judy Huang[9], Marco C. Pinho[10], Binu P. Thomas[10], Babu G. Welch[11,12], Denise C. Park[12], Vishal M. Patel[3,4], Argye E. Hillis[8], and Hanzhang Lu[1,2,6*]

[1]Department of Biomedical Engineering, Johns Hopkins University School of Medicine, Baltimore, MD, USA

[2]The Russell H. Morgan Department of Radiology and Radiological Science, Johns Hopkins University School of Medicine, Baltimore, MD, USA

[3]Department of Computer Science, Johns Hopkins University, Baltimore, MD, USA

[4]Department of Electrical and Computer Engineering, Johns Hopkins University, Baltimore, MD, USA

[5]Department of Diagnostic Radiology and Nuclear Medicine, University of Maryland School of Medicine, Baltimore, MD, USA

[6]F.M. Kirby Research Center for Functional Brain Imaging, Kennedy Krieger Institute, Baltimore, MD, USA

[7]Department of Biostatistics, Johns Hopkins Bloomberg School of Public Health, Baltimore, MD, USA

[8]Department of Neurology, Johns Hopkins University School of Medicine, Baltimore, MD, USA

[9]Department of Neurosurgery, Johns Hopkins University School of Medicine, Baltimore, MD, USA

[10]Department of Radiology, UT Southwestern Medical Center, Dallas, TX, USA

[11]Department of Neurologic Surgery, UT Southwestern Medical Center, Dallas, TX, USA

[12]Center for Vital Longevity, School of Behavioral and Brain Sciences, University of Texas at Dallas, Dallas, TX, USA

**\*Correspondence:** Hanzhang Lu, Ph.D., or Xirui Hou, Ph.D., Johns Hopkins School of Medicine, 360 Park Building, 600 N. Wolfe Street, Baltimore, MD, USA, 21287

Email: hanzhang.lu@jhu.edu or xhou4@jhu.edu

## Abstract

Cerebrovascular disease is a leading cause of death globally. Prevention and early intervention are known to be the most effective forms of its management. Non-invasive imaging methods hold great promises for early stratification, but at present lack the sensitivity for personalized prognosis. Resting-state functional magnetic resonance imaging (rs-fMRI), a powerful tool previously used for mapping neural activity, is available in most hospitals. Here we show that rs-fMRI can be used to map cerebral hemodynamic function and delineate impairment. By exploiting time variations in breathing pattern during rs-fMRI, deep learning enables reproducible mapping of cerebrovascular reactivity (CVR) and bolus arrive time (BAT) of the human brain using resting-state $CO_2$ fluctuations as a natural "contrast media". The deep-learning network was trained with CVR and BAT maps obtained with a reference method of $CO_2$-inhalation MRI, which included data from young and older healthy subjects and patients with Moyamoya disease and brain tumors. We demonstrate the performance of deep-learning cerebrovascular mapping in the detection of vascular abnormalities, evaluation of revascularization effects, and vascular alterations in normal aging. In addition, cerebrovascular maps obtained with the proposed method exhibited excellent reproducibility in both healthy volunteers and stroke patients. Deep-learning resting-state vascular imaging has the potential to become a useful tool in clinical cerebrovascular imaging.

# Introduction

Cerebrovascular diseases, such as acute ischemic stroke, atherosclerosis, Moyamoya disease, and vascular contributions to cognitive impairment and dementia (VCID), encompass a range of pathologies that affect different components of the cerebral vasculature and brain parenchyma. Additionally, brain tumors have also demonstrated altered vasculature which is key to their pathophysiology. Structural brain MR imaging, including T1, T2, diffusion-weighted image (DWI), susceptibility-weighted image (SWI), and magnetic resonance angiogram (MRA), is the current mainstay of imaging evaluation for these conditions. Advanced imaging such as perfusion and vessel wall imaging is also increasingly used in major medical centers.

Despite the progress, many of these conditions maintain high mortality and morbidity, and cost billions of dollars to the healthcare system[1]. Therefore, more advanced diagnostic and prognostic tools are urgently needed. Cerebrovascular reactivity (CVR) and bolus arrival time (BAT), which denote the brain vasculature's dilatory ability[2] and hemodynamic delay[3,4], respectively, represent two important markers of brain vascular function with proven utility in cerebrovascular conditions. For example, CVR has been suggested to be a sensitive biomarker in vascular cognitive impairment[5] and is currently undergoing multi-site clinical validation in the MarkVCID study[6]. BAT, sometimes presented in the forms of time-to-maximum (Tmax) and time-to-peak (TTP), is a promising biomarker in acute stroke and, when combined with DWI, can help delineate ischemic penumbra and guide triaging decisions in terms of recombinant tissue plasminogen activator (tPA) and/or endovascular thrombectomy[3,4,7-10].

Currently, CVR and BAT mappings are carried out using the administration of $CO_2$ enriched gas[2], vasodilatory pharmacological agents[11], or contrast agents[12,13]. The need to use exogenous agents in these measurements stems from the fact that these physiological parameters denote dynamic properties of brain vascular function. In the case of CVR, a vasoactive stimulus is needed to induce vessel dilation. To measure BAT, a tracer is needed to follow its path and timing along the cerebral vasculature. These methods, however, require additional procedures and equipment. Therefore, it is highly desirable to use imaging procedures comparable to standard anatomic MRI, e.g. acquired under resting-state, to assess advanced physiological parameters such as CVR and BAT. Under resting state, the arterial $CO_2$ concentration fluctuates as a result of spontaneous variations in breath-by-breath respiration. This presents an opportunity to use resting-state $CO_2$ change as an intrinsic marker to estimate CVR and BAT. Although several prior reports have demonstrated proof-of-principle studies (for CVR[14-18] and BAT[3,4,19-21] separately), these techniques generally suffered from low signal-to-noise ratio and variable image quality across patients, which was primarily due to the limited extent of natural variations in $CO_2$ during resting-state when compared to hypercapnia (HC) maneuvers with $CO_2$-inhalation.

In this study, we aimed to develop a robust, deep-learning framework to estimate CVR and BAT simultaneously from resting-state blood-oxygenation-level-dependent (BOLD) fMRI. The deep-learning network developed will work with fMRI data of any spatial and temporal resolutions as well as any number of time points. The network is applicable to data from healthy volunteers as well as patients with typical clinical vascular pathologies. Resting-state fMRI consists of a time series of 3D volumetric images representing the complex interplay of temporal fluctuations in both neural and vascular activities. Deep-learning networks have attracted much attention for their ability to harness high-dimension data and learn complex relationships through feature extraction and representation learning[22,23]. These methods have proven to be useful in applications across a wide range of disciplines in health care, such as breast cancer detection[24], heart disease identification[25-28], self-administration of medication[29], and personalized lab testing[30]. Here we employed a hierarchical deep-learning network to analyze resting-state BOLD images to extract CVR and BAT information. Training and testing of our deep-learning network included a wide range of cerebrovascular conditions to provide diverse data sources, including healthy volunteers, as well as patients with Moyamoya disease, brain tumor, and stroke, using $CO_2$-inhalation HC MRI data as labels, i.e., ground-truth. We also demonstrated clinical applications of this technique by comparing CVR and BAT with clinical variables. Furthermore, the reproducibility of the technique was evaluated in healthy volunteers and stroke patients.

## Results

**Fig. 1** shows the structure of our deep-learning framework. The inputs to the deep-learning network consisted of two parts. The primary input was the CVR and BAT maps obtained from the previous global-regression resting-state (GRRS) method[3,14]. Specifically, three 2D images, including GRRS CVR $\beta_0$, GRRS CVR $\beta_1$, and GRRS BAT, were used as the primary inputs. We used the parametric CVR and BAT maps instead of the raw BOLD image time series as inputs, so that our deep-learning network can be applied to BOLD data of any sample time points, repetition time (TR), or scan duration. These images had also been spatially normalized into Montreal Neurological Institute (MNI) standard space so that the network, once trained, can be applied to different field-of-views (FOVs), matrix sizes, and spatial resolutions. A supplementary input was also used in our deep-learning network. The supplementary input was based on the residual 4D image series after global-regression computation. We parcellated the whole brain into 133 regions-of-interest (ROIs)[31] and computed 133 2D cross-correlation (CC) maps, in each of which the residual time course of one ROI was used as the reference time course for voxel-wise CC calculation. This additional input accounts for residual vascular information and regional variations in vascular responses that are present in the BOLD data but not captured in the global-regression results[32-34].

The outputs of the network were the estimated 2D images of deep-learning resting-state CVR (DLRS CVR) and deep-learning resting-state BAT (DLRS BAT) maps. Between inputs and outputs, the architecture of the deep-learning network consisted of an encoder module and a decoder module. The encoder module contained a primary encoder that extracted vascular features from the primary inputs, i.e., GRRS CVR and GRRS BAT, and a supplementary encoder that was applied to the supplementary inputs, i.e., the CC maps from the residual BOLD data. The decoder module contained a CVR-specific component and BAT-specific component. Each component integrated the latent representations in the primary and supplementary encoders, and provided an estimation of CVR (or BAT) map. More details of our deep-learning network architecture and training process are described in Methods.

The datasets used in our study are summarized in **Table 1** and detailed in **Supplementary Table 1-4**. The deep-learning network was trained and tested on datasets from 232 participants, each of whom underwent a resting-state fMRI and an HC MRI scan. We performed $K$-fold cross-validation ($K = 5$) to evaluate our model[29]. That is, the datasets were randomly divided into five subgroups. For each fold, a single subgroup was retained as test data, whilst all other subgroups collectively were used for training. This process was repeated five times, with each of the five subgroups used exactly once as the test data.

**Fig. 2a,b,c** show representative images of DLRS CVR/BAT for healthy, Moyamoya disease, and brain tumor, together with GRRS and ground-truth HC CVR/BAT images. Visual inspection suggested that the deep-learning images resembled the ground-truth images, and were superior to the global-regression maps. Quantitative evaluations were based on Pearson

cross-correlation, structure similarity index measure (SSIM), peak signal-to-noise ratio (PSNR), and root-mean-square error (RMSE) between the RS based CVR/BAT maps with the ground-truth HC CVR/BAT, as shown in **Fig. 2d-k**. In all quantitative indices evaluated, the deep-learning results revealed a significantly higher congruency with the HC results, when compared to those from the global-regression approaches[4,14].

We conducted an ablation study to demonstrate the efficacy of the proposed network architecture, particularly the necessity of the supplementary encoder. As shown in **Extended Data Fig. 1**, the DLRS CVR and DLRS BAT maps in Moyamoya patients revealed a lower spatial correlation with the ground-truth maps if the supplementary encoder is omitted, although in the healthy participants, the results did not reveal a difference.

Next, we sought to demonstrate the sensitivity of the CVR and BAT maps in detecting vascular abnormalities and treatment effects. Region-of-interest (ROI) values were compared between ipsilateral and contralateral regions in Moyamoya, brain tumor, and stroke patients (**Fig. 3a,b,c**). For Moyamoya patients, we examined the DLRS CVR/BAT differences in the middle cerebral artery (MCA) territories between the hemispheres that underwent revascularization surgery and those that did not. We focused on MCA territories because revascularization procedures typically aim to recover perfusion in these regions. In the brain tumor and stroke patients, the comparisons were primarily focused on the lesion regions versus the contralateral normal regions. Quantitative results of these comparisons are summarized in **Fig. 3d-i**. As a reference, the HC CVR/BAT values revealed significant differences between ipsilateral and contralateral regions for all comparisons conducted. The diseased side showed a lower CVR and longer BAT. From the DLRS data, we observed a significant difference between ipsilateral and contralateral regions in all comparisons conducted, the directions of which were consistent with those in the HC data. The GRRS results also revealed laterality-related differences, although the effect sizes were smaller than that of DLRS in comparisons.

We further investigated whether the imaging data, specifically CVR and BAT, were correlated with clinical variables. We examined the associations between CVR/BAT and arterial stenosis grades for Moyamoya disease and with tumor grades in brain tumor patients. As shown in **Extended Data Fig. 2**, DLRS CVR and DLRS BAT results revealed a significant correlation with the clinical variables, and the correlation coefficient values were greater than that for GRRS results.

We also studied age-related differences in DLRS CVR/BAT in the healthy participants, and compared them to those from ground-truth HC CVR/BAT. Consistent with previous studies[35], we selected the occipital lobe as a reference, which was thought to be most age-preserved in the brain[36,37], and normalized all other brain regions against the occipital CVR/BAT. As shown in **Extended Data Fig. 3**, DLRS CVR revealed significant decreases with age across the majority of brain regions, while DLRS BAT increased with age (FDR-adjusted $p < 0.05$). The Dice coefficients between DLRS CVR and HC CVR were 0.78 for age-decrease effects, whereas the Dice coefficients when using the global-regression approach were 0.08. Similarly, for BAT, the Dice coefficients were 0.77 and 0.04, respectively.

To conduct a test-retest reproducibility assessment, in a new dataset of healthy participants (N=67) and stroke patients (N=30), we performed two identical resting-state fMRI scans in the same session. **Fig. 4a** displays the DLRS CVR/BAT from both scans on a healthy participant, along with GRRS and HC maps. Note that the HC scan was only performed once. As shown in **Fig. 4c,d**, the DLRS CVR/BAT images in healthy subjects consistently revealed a significantly higher correspondence with HC CVR (i.e., spatial Pearson cross-correlation) than those from GRRS CVR/BAT ($p < 1.0 \times 10^{-5}$ for all tests). **Fig. 4e,f,i,j** showed the scatter plots of DLRS CVR and DLRS BAT obtained from two scans in healthy participants, together with those from the GRRS approach. We observed that the deep-learning results were distributed closer to the unity line, with a smaller difference between the two scans. The intraclass correlation coefficients (ICC) of the DLRS CVR and the GRRS CVR were 0.863 (95% CI, 0.857-0.868) and 0.627 (95% CI, 0.615-0.640), respectively. Similarly, the ICCs of the DLRS BAT and the GRRS BAT were 0.864 (95% CI, 0.859-0.870) and 0.386 (95% CI, 0.368-0.404), respectively. The ICC analysis revealed that the deep-learning approaches showed a better agreement between two scans in both CVR and BAT images of healthy participants ($p < 1 \times 10^{-5}$ for CVR and BAT). **Fig. 4b** depicted our DLRS CVR/BAT images with reference to DWI and T2-weighted images for a stroke case. In the stroke datasets, the scatter plots between two scans were consistent with those from healthy participants (**Fig. 4 g,h,k,l**), indicating a smaller difference between two scans from deep-learning results. The ICCs of our DLRS CVR and DLRS BAT were 0.874 (95% CI, 0.867-0.881) and 0.857 (95% CI, 0.848-0.865), respectively, again with significant improvements ($p < 1 \times 10^{-5}$ for CVR and BAT) compared with those from the previous approach (GRRS CVR: 0.724 (95% CI, 0.710-0.739); GRRS BAT: 0.561 (95% CI, 0.539-0.581)). The ICCs from stroke datasets were higher than from healthy datasets due to the longer scan time and no repositioning between two scans.

**Discussion**

The past few years have witnessed the advent of several promising applications of deep-learning networks in resting-state fMRI analyses[38-42]. These studies were mainly concerned with measuring neural signals from fMRI and developing new biomarkers based on neural signature of the brain. A major novelty of the present work is that our study primarily focuses on the fMRI vascular signals instead and uses them to generate 3D maps of cerebrovascular reactivity and bolus arrival time. Here we described a deep-learning approach to reconstructing brain CVR and BAT maps from resting-state BOLD images. The basis of our approach is to exploit arterial $CO_2$ fluctuations induced by breath-to-breath respiratory variations as an intrinsic "contrast agent" to map cerebrovascular physiology. The encoder-decoder framework used in the present study is analogous to the U-shape networks for image quality enhancement in low-dose CT[43], PET[44,45], and MRI[46]. It allows the deep-learning model to use the results of existing non-DL methods as inputs (i.e., GRRS CVR and GRRS BAT), but improves the quality of the image by taking advantage of the prior knowledge gained during model training. Furthermore, we included a supplementary encoder module in the network to take the residual BOLD signal into consideration. This residual signal was discarded in previous global regression methods, but may contain region-specific vascular information associated with vasodilatory response function or local properties of neurovasculature[32,33]. Moreover, the global BOLD signal used in the previous regression methods may contain non-vascular-origin fluctuations (e.g., global variations in neural activity due to vigilance shifting[47,48]), which could be alleviated by deep-learning method because vascular-based HC CVR/BAT was used as label in training.

The present study showed that all quantitative metrics of CVR and BAT from our model revealed a more consistent performance when compared to previous approaches across several medical conditions. To better understand the reason for this improvement, we further evaluated the contribution of each component in our network through an ablation study. We found that the performance without the supplementary encoder deteriorated compared to the full network, especially among patients with cerebrovascular pathology. This may be attributed to the hemodynamic abnormalities in the lesion regions, resulting in heterogeneous signals across the brain and making the global regression BOLD signal less informative[49]. Therefore, integrating the information from residual signal with the global-regression results enables the full model to outperform the sub-component models.

This study further demonstrated the sensitivity of the deep-learning derived hemodynamic maps in detecting vascular abnormalities for various brain diseases. Due to the blockage of cerebral blood vessels, Moyamoya disease and ischemic stroke will cause a reduction in the cerebrovascular reserve and a delay in blood arrival[3,50-52]. Our data suggested that the deep-learning derived maps can successfully delineate regions with deteriorated CVR and delayed BAT. Similarly, the deep-learning maps also identified cerebrovascular abnormalities in brain tumors. It is known that neovasculature formed in and around high-grade primary brain tumors due to angiogenesis is immature and not equipped with vascular smooth muscles[53]. Thus, absent or diminished regional CVR is consistent with this aspect of tumor biology[54,55].

We also provided some early evidence that the deep-learning derived hemodynamic maps may be useful in patient triaging or evaluating treatment effectiveness. In Moyamoya patients, the DLRS CVR/BAT metrics in those who underwent revascularization surgery were significantly improved compared to those who did not[56]. Moreover, our results indicated that the cerebrovascular dilation ability deteriorated with worse stenosis grades from MR angiography. For brain tumor patients, we observed that CVR/BAT was correlated with tumor WHO grades. These findings suggest that deep-learning derived hemodynamic maps have the potential to be further developed into novel biomarkers for disease diagnosis and treatment monitoring.

Additionally, we examined the age dependence of DLRS CVR/BAT in a lifespan cohort. We found that most regional CVR from our model tended to decrease while regional BAT increased with age. Our result demonstrated more consistency with the $CO_2$-challenged study than from global-regression approaches. Our results are also in line with the notion that vascular response to $CO_2$ challenge becomes diminished as we age[57-60].

To ensure the applicability of our deep-learning network to both healthy participants and patients with diseases, we trained our model using datasets from participants with a range of medical conditions as well as healthy volunteers. Besides, our deep-learning method converted the raw fMRI signals into cross-correlation maps before the data were used for training and testing. An advantage of this approach is that the trained network can be used for any resting-state fMRI datasets, regardless of the TR or time points used in the acquisition.

Furthermore, we conducted an independent test-retest study using a separate dataset. Our results demonstrated a high consistency in CVR and BAT maps using the proposed method when compared with previous approaches.

While the present study primarily investigated the clinical utility of the proposed hemodynamic mapping method, we would like to note that our method can also be combined with conventional fMRI analysis approaches to obtain a better interpretation of neural signals. BOLD fMRI is long known to be an indirect assessment of neural activity and is influenced by the microvascular function of the brain[32,33,61]. In fact, the presence of a BOLD fMRI signal is dependent on the vasodilation associated with neural activity[62]. Therefore, the CVR map obtained from the present method can be used to normalize or calibrate the functional connectivity results that are commonly used in the literature. Importantly, both CVR and functional connectivity maps can be estimated from the same data without any additional data acquisition, and the two maps are automatically coregistered. Therefore, our deep-learning method may offer a promising approach for vascular-corrected fMRI quantification in future studies.

Although our deep-learning method provides a significant improvement over the previous global regression method, it also has several limitations. First, the CVR and BAT obtained from our deep-learning model were in relative units, rather than in absolute units of percentage per millimeter mercury (mmHg) of $CO_2$ change. Hence, our approach is more suited for diseases in which CVR and BAT deficits are regional. Additionally, the current deep-learning results presented in stroke patients have not been validated with the ground-truth hypercapnic method, due to practical challenges in performing $CO_2$ inhalation in this group of patients. Future studies are needed to validate its clinical utility.

In summary, our study shows that cerebrovascular reactivity (CVR) and bolus arrival time (BAT) mappings using deep-learning model from resting-state functional MRI provided a task-free approach to assess cerebrovascular dilation ability and arterial delivery time across the brain. This technique demonstrated excellent performance when applied to healthy participants across the lifespan, and in patients with stroke, Moyamoya disease, or brain tumor. CVR and BAT mapping with resting-state fMRI may provide a new platform for developing physiological biomarkers in brain diseases.

## Methods

Study participants

Participants were recruited from two sites: the University of Texas Southwestern Medical Center (UTSW), Dallas, TX; Johns Hopkins University (JHU), Baltimore, MD. The study and procedures were approved by the Institutional Review Boards of UTSW and JHU, in compliance with all ethical regulations. **Table 1** and **Supplementary Table 1-4** summarize participant characteristics and imaging parameters. Specifically, 236 healthy participants and 34 Moyamoya patients were recruited at UTSW site[36,63]. 15 Moyamoya patients, 14 brain tumor patients, and 68 stroke patients were recruited at JHU site. The Moyamoya patients were characterized by severe stenosis/occlusion predominantly involving the intracranial segments of the internal carotid arteries, diagnosed between 2014 and 2019 (**Supplementary Table 1**). The 14 de novo brain tumor subjects were recruited between 2016 to 2019 before surgical operation (**Supplementary Table 2**). The 68 stroke subjects were enrolled between 2012 and 2019[64,65] (**Supplementary Table 3 and 4**). The stroke subject selection criteria were: (a) clinically confirmed stroke within 16 months prior to the MRI scan; (b) at least a T2-weighted image or diffusion-weighted image (DWI) in the same session of the resting-state fMRI scan.

Each participant underwent one resting-state fMRI scan. A subset of 67 healthy participants (from the UTSW site) and 30 stroke patients (from the JHU site) also underwent a second resting-state fMRI scan in the same session for reproducibility assessment (**Table 1** and **Supplementary Table 4**). The second resting-state fMRI scan on healthy participants was performed with a break and repositioning. The stroke patients did not undergo the hypercapnic CVR scan, whereas all the other participants received a hypercapnic CVR scan.

MRI protocols

All MRI examinations were performed on 3T MRI scanners (Achieva, Philips Medical Systems, Best, The Netherlands). Each participant received a T1-MPRAGE scan and a resting-state fMRI scan. The resting-state scan was performed while the subject was asked to lie still without performing any task. Except for stroke patients, each participant also underwent the hypercapnic scan, during which 5% $CO_2$ was used as a vasodilative stimulus. The details of the hypercapnic scan have been previously described[66,67]. Other relevant sequences performed on the participants are listed in **Table 1**.

Resting-state BOLD Processing

For resting-state BOLD processing, we used Statistical Parametric Mapping (SPM12, University College London) and in-house Matlab (version 2019a, MathWorks) scripts. The processing pipeline is illustrated in **Extended Data Fig. 4**. Briefly, the resting-state BOLD images first underwent standard preprocessing steps, including motion correction, slice timing correction, normalization to Montreal Neurological Institute (MNI) standard brain space via MPRAGE image, and spatial smoothing using a Gaussian filter with a full-width half-maximum of 8 mm. The BOLD image series were detrended and band-pass filtered with a frequency of [0 Hz, 0.1164 Hz][14].

After applying the band-pass filter, a general linear regression analysis was performed using the cerebellum BOLD signal as the reference signal time course (i.e., independent variable) and the voxel-wise BOLD signal as dependent variable[14], with 12 motion vectors as covariates (6 band-pass filtered motion parameters and their squares)[68,69], yielding GRRS CVR coefficient maps (i.e., GRRS CVR $\beta_0$, GRRS CVR $\beta_1$)[66]. We used the cerebellum signal, instead of the whole-brain global signal, as the reference signal because cerebellum territories are typically unaffected in the studied patient populations, whereas the global signal can be compromised[14,50]. We then conducted feature scaling on the GRRS CVR coefficients by converting the voxel-wise coefficients into Z-scores, which were used as a primary input in the deep-learning network. To obtain the GRRS CVR value for comparison with the DLRS CVR results, we calculated the ratio between the coefficients, i.e., $\beta_1/\beta_0$, and converted the map to Z-score[14].

For GRRS BAT map, the BOLD time course of each voxel within the brain was extracted and then shifted between ±9 s with an increment of 0.1 s. The search range of ±9 s is based on the previous literature[19,70-72]. A general linear model was used for each shifted time course with cerebellum time course as the independent variable and 12 motion vectors as covariates (6 band-pass filtered motion parameters and their squares)[68]. The optimal shift at each voxel was identified as the value that maximizes the full model's coefficient of determination ($R^2$) to account for possible collinearities between regressors[72]. The optimal shift for each voxel then underwent feature scaling, i.e., normalization to Z-score, yielding the GRRS BAT map, which was used as another primary input to the DL model.

To obtain supplementary inputs for the DL model, the residual BOLD signal after global regression was parcellated into 133 regions-of-interest (ROIs), based on the Neuromorphometrics Atlas in standard SPM12 without extensions[31]. Then, using the spatially averaged residual signal time-course in each ROI as a reference, cross-correlation maps were calculated. The cross-correlation maps were then normalized to Z-scores and yielded 133 supplementary inputs.

Hypercapnic BOLD Processing

The processing pipeline for the hypercapnic BOLD is illustrated in **Extended Data Fig. 5**. The preprocessing pipeline was similar to that of the resting-state BOLD, which consisted of motion correction, slice timing correction, normalization to MNI standard brain space, and spatial smoothed using a Gaussian filter with a full-width half-maximum of 8 mm. The $EtCO_2$ time course was temporally aligned with the reference BOLD time course (i.e., cerebellum BOLD time course) to account for the time it takes for $CO_2$ to travel from the lung (where the $EtCO_2$ was recorded) to the brain (where the BOLD signal was recorded). A general linear model was performed using each voxel BOLD signal as the dependent variable, $EtCO_2$ as the independent variable, and linear drift term as the covariate, yielding HC CVR coefficient maps (i.e., HC CVR $\beta_0$, HC CVR $\beta_1$). CVR was then computed as $CVR = \frac{\beta_1}{\beta_0 + bEtCO_2 \times \beta_1}$. We noted that HC CVR was not calculated as $\frac{\beta_1}{\beta_0}$, but instead contained the $bEtCO_2 \times \beta_1$ term, so that the measured HC CVR was in reference to basal $EtCO_2$ state under room air[66]. The HC CVR was then converted to Z-score map.

The HC BAT was quantified as the time delay between the synchronized $EtCO_2$ time course and the voxel-wise BOLD signal time course. The analysis was performed by shifting the voxel-wise time course in the range of [-10 s, 30 s] with an increment of 0.1 s, based on the previous literatures[2]. A generalized linear analysis was performed for each shifted time course with synchronized $EtCO_2$ time course as the independent variable and linear drift term as the covariate. The optimal shift was identified similar to that in the resting-state BAT method[72]. The optimal shift for each voxel then underwent Z-score normalization, yielding the HC BAT map. These HC CVR and BAT maps are used as labels in the training of the DL network.

Image Padding and Clipping

The normalized images were in MNI space with a dimension of 91×109×91. We resized the image to 96×112×91 by padding zero voxels on the border of x, y directions, ensuring the input and label 2D image stacks were compatible with our encoder-decoder framework. The ground-truth HC CVR and HC BAT maps were clipped to the range of [-5, 5] to keep the image in a reasonable scale. **Extended Data Fig. 6** showed that only 0.16% voxels in HC CVR and 0.00008% voxels in HC BAT images clipped to the maximum or minimum intensities.

Encoder-decoder framework

Our proposed DL framework is based on the "auto-encoder network" design and includes encoders and decoders, as shown in **Fig. 1** and **Extended Data Table 1**. In an auto-encoder model[43,44,73], the encoder module converts high-dimensional data into embedded representations whereas the decoder module reconstructs high-dimensional data. In our model, our network contains two encoders. The primary encoder processed input images of GRRS CVR $\beta_0$, RS GRCVR $\beta_1$, and GRRS BAT. The supplementary encoder processed the 133 correlation maps. The primary and supplementary encoders contained an identical network architecture similar to the contracting path in U-Net[74]. Within each encoder, there were five convolutional blocks. Each block consisted of the following sequential layers: convolutional layer, rectified linear unit (ReLU) layer, batch normalization layer, convolutional layer, ReLU layer, batch normalization layer, and max pooling layer. In each block, we doubled the number of feature channels, while we cut the spatial dimensions in half. The specific configuration of each block is listed in **Extended Data Table 1**.

The decoder module contained two identical decoders. The decoders utilize high-dimensional representation provided by the encoders and perform customized synthesis for output maps, in our case CVR and BAT (**Fig. 1**)[75]. Each decoder consisted of four deconvolutional blocks based on the expansive path in U-Net[74]. Each block contained the following sequential layers: transpose convolutional layer, concatenation layer, convolutional layer, ReLU layer, batch normalization layer, convolutional layer, ReLU layer, and batch normalization layer. At the end of the decoder, the resulting feature map passed through 1×1 convolutional layer and 5×tanh activation unit layer to generate CVR or BAT in the range of [-5, 5], respectively. The specific configuration of each block is listed in **Extended Data Table 1**.

Training and testing of the DL network

The DL network was trained using GRRS CVR $\beta_0$, GRRS CVR $\beta_1$, GRRS BAT, and cross-correlation maps as input images, and HC CVR and HC BAT as ground-truth images. We defined the loss function as the L1-norm error between the prediction and ground-truth images. The DL network was implemented by using the PyTorch library. The AdaBelief was used as the optimizer[76] to minimize the loss function and update the network parameters iteratively through back-propagation. A learning rate of $5\times10^{-5}$, epsilon of $1\times10^{-12}$, and a batch size of 64 were used in our training for 100 epochs. Data augmentation, including horizontal flipping and vertical flipping, was applied to the Moyamoya and brain tumor datasets to increase the size of our training data and thus reduce overfitting[77]. The final weights of the network were determined based on training results using all 232 datasets. For the purpose of testing, fivefold cross-validation was used[29,44,78]. Specifically, the 232 resting-state/hypercapnic datasets were divided into five subgroups, each consisting of similar numbers of healthy volunteers and patients. For the k-th fold (k=1,2,…,5), the DL network was trained based on data from the four other subgroups and then tested on the k-th subgroup. We trained the network using two Nvidia Titan RTX graphics processing units (duration typically around 24h). During testing, the typical inference time for one testing sample of a participant is around 0.5s.

We conducted an ablation study to further investigate the necessity of the supplementary encoder in the DL network. The inputs associated with the supplementary encoder, i.e., 133 cross-correlations maps, were removed. We compared the CVR and BAT maps obtained from the ablated model to those from the full model in terms of their correlations with the HC maps.

Quantitative assessment of the DL results

To compare the DL-derived maps to ground-truth hypercapnic maps, we computed four metrics: 1) spatial Pearson cross-correlation between DLRS and HC maps; 2) Structural similarity index measure (SSIM), which aims to account for multiple factors used in human visual perception and integrates similarities of two images in terms of luminance, contrast, and structure.; 3) Peak signal-to-noise ratio (PSNR) which is defined as the ratio between the maximum signal in the DLRS image and mean square error of the voxel-wise difference between DLRS and HC images; 4) root mean square error (RMSE) which quantifies the voxel-wise L2-norm error between two images. In general, a higher cross-correlation, SSIM, and PSNR or a lower RMSE indicates a better prediction closer to the ground-truth images.

The sensitivity of DLRS CVR and DLRS BAT images to brain pathologies was examined in three patient cohorts: Moyamoya disease patients, stroke patients, and brain tumor patients. In Moyamoya patients, we aimed to evaluate whether DLRS CVR (or DLRS BAT) in affected hemispheres that have received revascularization surgery is different from those that have not. Of the patients we have studied, 75 hemispheres suffered from stenosis based on MRA. Of these, 30 have had revascularization surgery at the time of the MRI scan. The remaining 45 have not had surgery. We compared DLRS CVR and DLRS BAT values between hemispheres with and without revascularization surgery. Regional CVR and BAT values were obtained from the perfusion territories of the middle cerebral artery (MCA) based on a perfusion atlas[79]. We focused on MCA territories for Moyamoya disease patients because revascularization procedures typically aim to recover perfusion in these regions.

The grade of MCAs stenosis of each participant with Moyamoya disease was rated independently by a neuroradiologist (M. P., with >10 years of clinical experience) who was blinded to the CVR/BAT results. The rating was made by using a previously published angiographic scoring system adapted to MR angiography[80]. The association between DLRS CVR (or DLRS BAT) and MCA stenosis grade was evaluated by using linear regression in hemispheres which did not undergo revascularization surgery.

We then analyzed data from patients with brain tumors to compare DLRS CVR and DLRS BAT values between lesion and contralateral normal regions. The lesion regions were delineated on T2-FLAIR images by a rater (X. H., with >5 years of experience and verified by D. L., with >10 years of clinical experience) blinded to the CVR and BAT maps. Control regions were obtained by mirror-flipping the lesion ROI with regard to the mid-line of the image. The tumor grade is based on the 2016 World Health Organization (WHO) updated criteria of brain tumors[81].

For the stroke data, similar approaches were used to obtain manually defined ROIs. Since some of the stroke patients were scanned in acute/subacute phase while others were scanned in a chronic phase, different types of anatomic images were used for the ROI drawing. For acute/subacute stroke patients, the lesion regions were manually delineated on DWI images.

The T2-weighted images were used for ROIs drawing on lesion regions for chronic stroke patients.

Reproducibility Study

We further assessed the reproducibility of our proposed method on an independent dataset that contains 67 healthy participants and 30 stroke patients. Each of the 67 healthy participants underwent two resting-state fMRI scans, with a break and repositioning in-between. A hypercapnic CVR scan was also performed. For the 30 stroke patients, two resting-state fMRI scans were performed in the same session (without repositioning). The DLRS CVR and DLRS BAT maps were parcellated into 133 ROIs, based on the Neuromorphometrics Atlas in SPM12[31]. Pearson cross-correlation and intraclass correlation (ICC) between the two maps were computed.

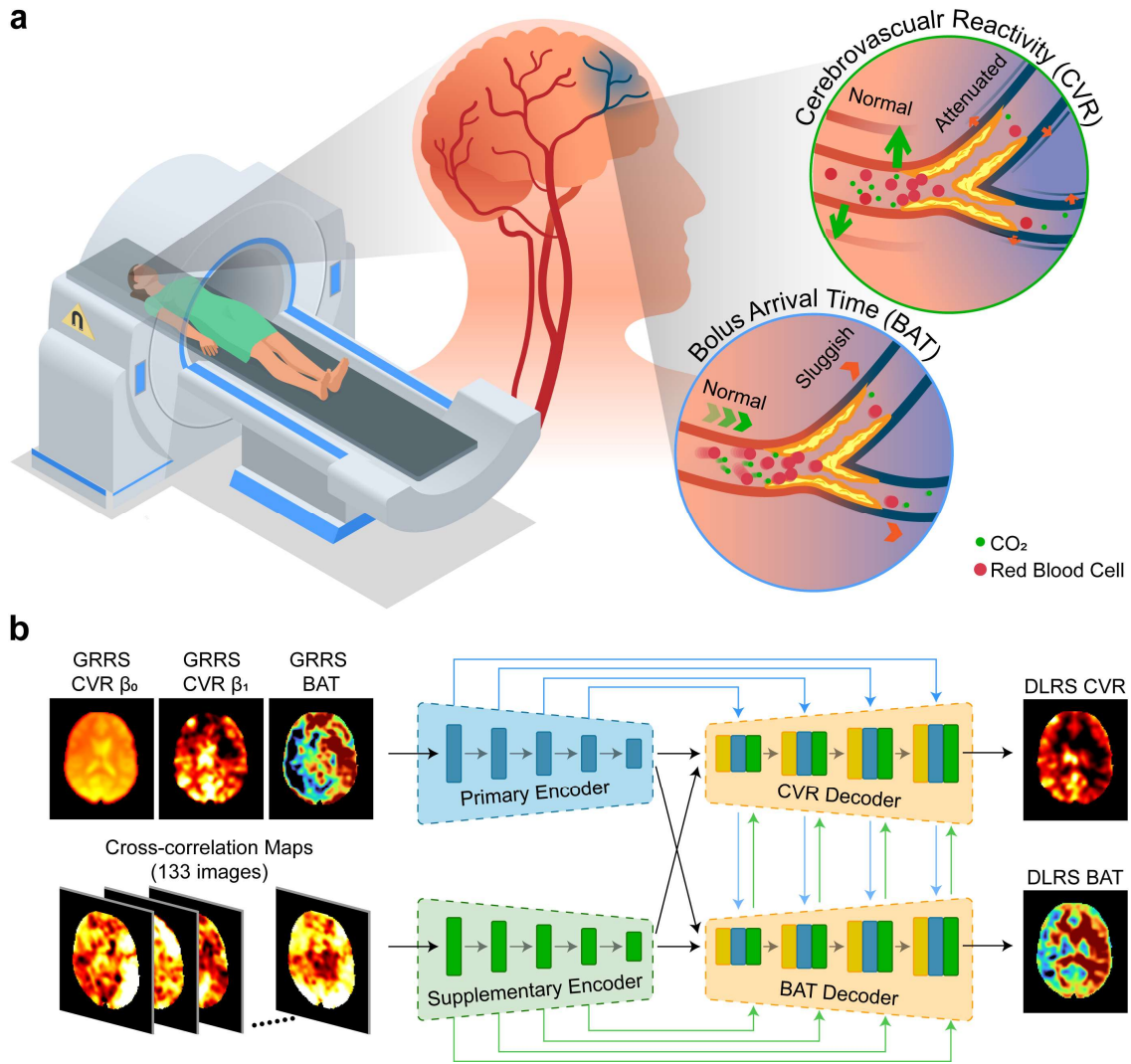

**Fig. 1 | Overview of MRI experiment and deep-learning network used in this work. a**, An illustration of MRI experiment to measure brain hemodynamic function. Spontaneous fluctuations in breathing pattern during resting-state MRI result in changes in $CO_2$ level in the arterial blood. This $CO_2$ change can be used as an intrinsic marker for the estimation of cerebrovascular reactivity (CVR) and bolus arrival time (BAT) using deep-learning network. **b**, Architecture of the deep-learning network. An encoder-decoder network was used, where primary and supplementary features of the image series were analyzed, and then fused to generate the outcome measures of resting-state CVR and BAT maps.

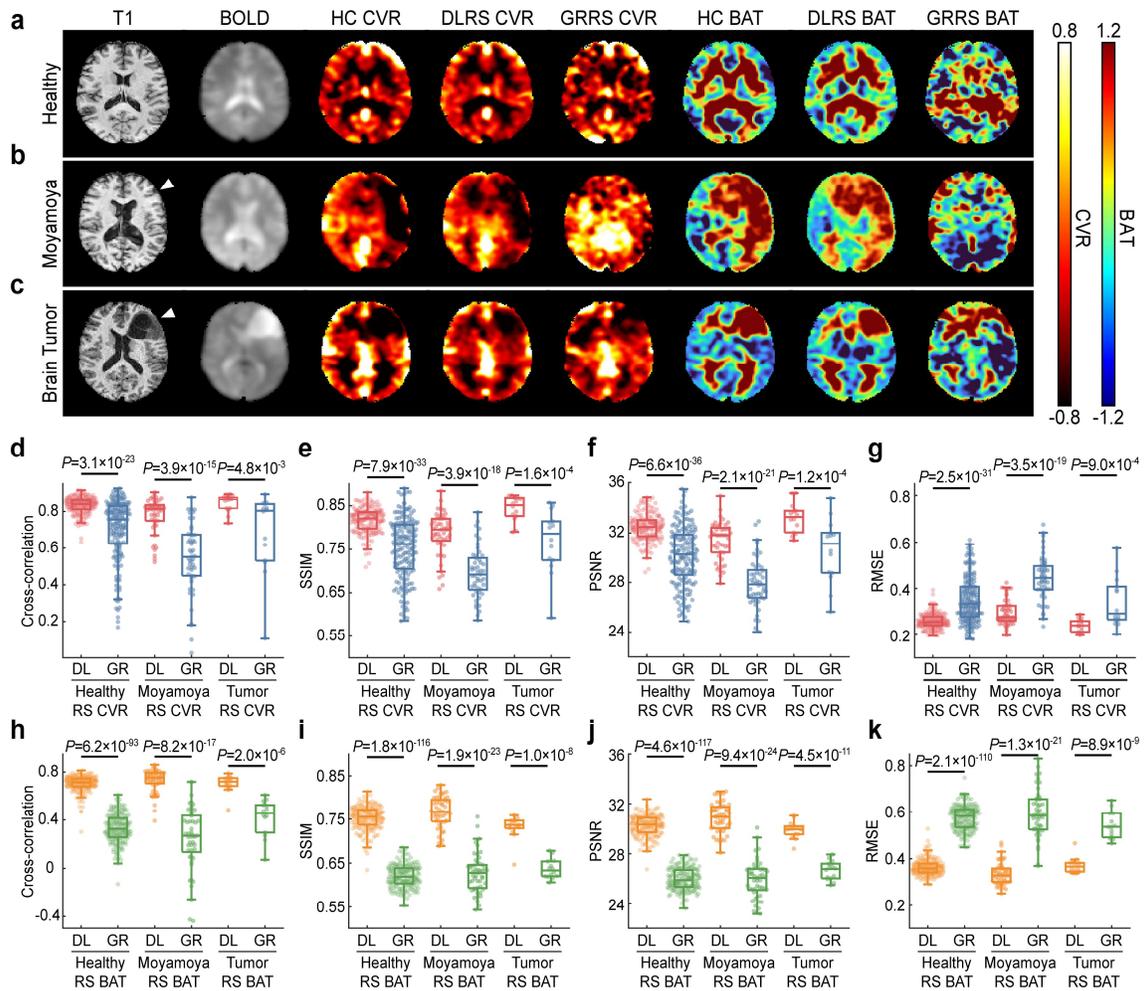

**Fig. 2 | Representative images and quantitative results of the deep-learning resting-state cerebrovascular reactivity (DLRS CVR) and bolus arrival time (DLRS BAT). a-c,** Representative images from a healthy volunteer **(a)**, Moyamoya disease patient **(b)**, brain tumor patient **(c)**. From left to right, the images are T1-weighted anatomic images, raw BOLD images, hypercapnic (HC) CVR, DLRS CVR, global-regression resting-state (GRRS) CVR, HC BAT, DLRS BAT and GRRS BAT. **d-g,** Quantitative indices indicating similarity between resting-state CVR maps and ground-truth HC CVR maps. Two types of resting-state CVR maps were studied: the proposed DLRS CVR and an existing GRRS CVR. Four similarity indices were studied, including Pearson cross-correlation **(d)**, structure similarity index metric (SSIM) **(e)**, peak signal-to-noise ratio (PSNR) **(f)**, root-mean-square error (RMSE) **(g)**. **h-k**, Quantitative indices indicating similarity between resting-state BAT maps and ground-truth HC BAT maps.

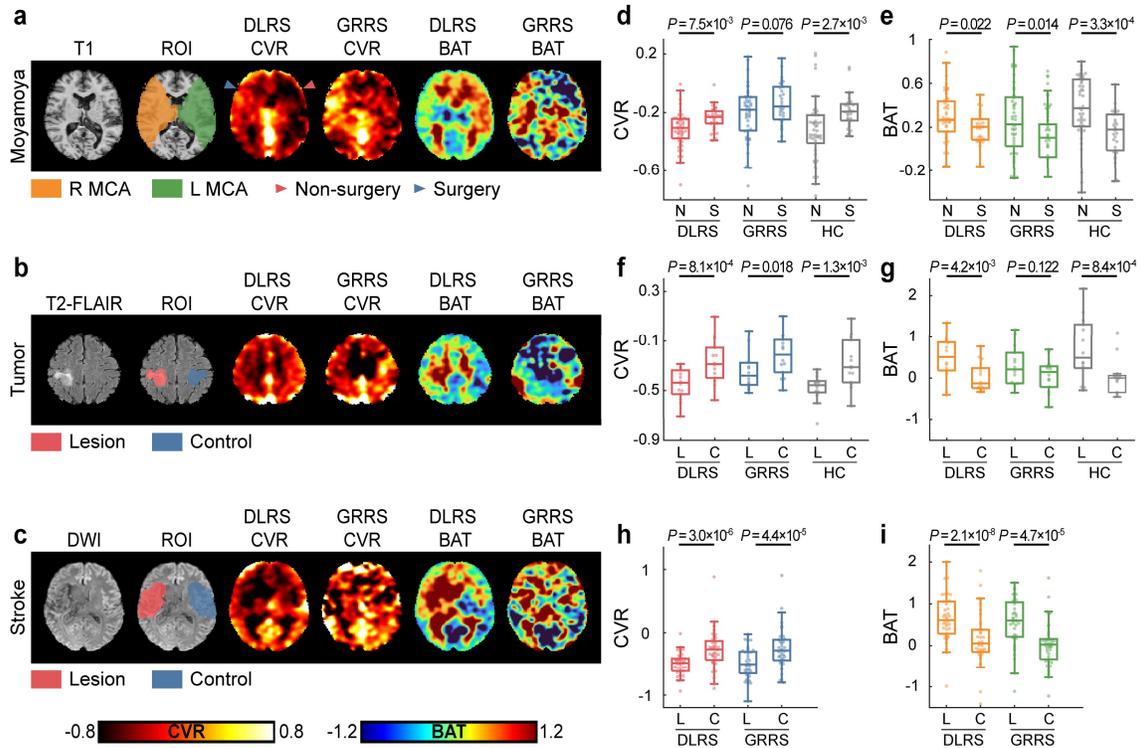

**Fig. 3 | The performance of deep-learning resting-state cerebrovascular reactivity (DLRS CVR) and bolus arrival time (DLRS BAT) in detecting brain pathologies. a,** A patient with Moyamoya disease who suffered from bilateral stenosis with the right hemisphere undergoing a revascularization surgery. Lower CVR and longer BAT can be seen in the non-surgical hemisphere. From left to right, the images are T1-weighted image, the middle cerebral artery (MCA) perfusion ROIs, DLRS CVR, global-regression resting-state cerebrovascular reactivity (GRRS) CVR, DLRS BAT and GRRS BAT. **b,** A diffuse astrocytoma patient with T2-FLAIR image, the lesion/control ROIs, DLRS CVR, GRRS CVR, DLRS BAT and GRRS BAT. **c,** A stroke patient with diffusion-weighted image (DWI) image, the lesion/control ROIs, DLRS CVR, GRRS CVR, DLRS BAT and GRRS BAT. **d,e,** Summary of CVR and BAT data in Moyamoya patients, when comparing their values between the surgically revascularized (S) hemispheres and the non-surgery (N) hemispheres. The effect sizes of two groups of DLRS CVR, GRRS CVR and HC CVR were 0.65, 0.42, and 0.73, respectively. The effect size of DLRS BAT, GRRS BAT and HC BAT were -0.55, -0.59, and -0.89. **f,g,** Summary of CVR and BAT data in brain tumor patients, when comparing between lesion (L) and contralateral control (C) areas. Tumor regions revealed a lower CVR and a longer BAT. The effect sizes of two groups of DLRS CVR, GRRS CVR, HC CVR, DLRS BAT, GRRS BAT, and HC BAT were 1.16, 0.72, 1.10, -0.93, -0.44, and -1.15, respectively. As can be seen, DLRS parameters showed a larger effect size than the existing GRRS method. **h,i,** Summary of CVR and BAT data in stroke patients, when comparing values between lesion (L) and contralateral control (C) areas. The effect sizes of group comparisons were 0.89, 0.75, -1.15, and -0.75 for DLRS CVR, GRRS CVR, DLRS BAT and GRRS BAT, respectively. DLRS parameters generally showed a larger effect size than the GRRS method.

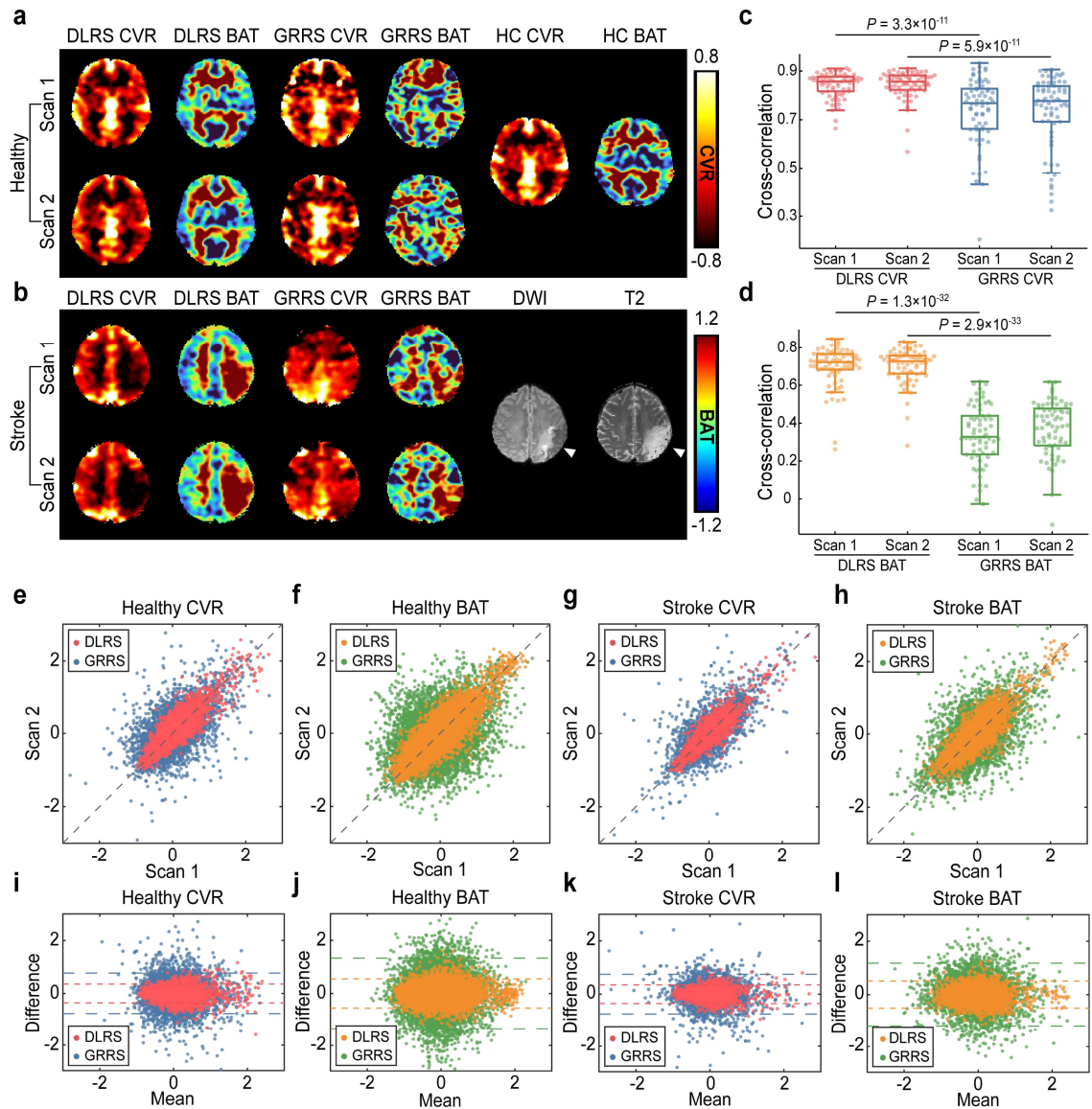

**Fig. 4 | Reproducibility of deep-learning resting-state cerebrovascular reactivity (DLRS CVR) and bolus arrival time (DLRS BAT). a,** A test-retest example from a healthy participant. The participant underwent two RS-MRI runs in the same session. From left to right, DLRS CVR, and DLRS BAT, global-regression resting-state cerebrovascular reactivity (GRRS) CVR, and GRRS BAT, hypercapnic (HC) CVR and HC BAT. **b,** A test-retest example from a stroke participant. The patient underwent two RS-MRI runs in two sessions. From left to right, DLRS CVR, DLRS BAT, GRRS CVR, GRRS BAT, diffusion-weighted image (DWI) and T2-weighted image. **c,** Pearson cross-correlation coefficient between DLRS CVR and HC CVR, together with those between GRRS CVR and HC CVR. **d,** Pearson cross-correlation coefficient between DLRS BAT and HC BAT, together with those between GRRS BAT and HC BAT. **e,f,** Scatter plots between two repeated scans for CVR and BAT across 133 ROIs in healthy participants. Each plot displayed data from both DLRS and GRRS methods. **g,h,** Scatter plots between two repeated scans for CVR and BAT in stroke patients. **i-l,** Bland-Altman plots of the CVR and BAT results in healthy participants and stroke patients.

**Table 1 | Demographics and MRI sequence parameters of datasets used in this work.**

| | Training and Test (5-fold cross validation) | | | Additional Clinical Test | Reproducibility Test | |
|---|---|---|---|---|---|---|
| | Healthy | Moyamoya | Brain Tumor | Stroke | Healthy | Stroke |
| N | 169 | 49 | 14 | 38 | 67 | 30 |
| Age, yr (mean ± s.d. (range)) | 51±20 (20-88) | 41±12 (18-72) | 42±18 (21-81) | 55±13 (27-87) | 52±18 (24-90) | 57±13 (24-80) |
| Female | 104 | 43 | 5 | 15 | 42 | 11 |
| rs-fMRI Sequence Parameters | | | | | | |
|   No. Scans | 1 | 1 | 1 | 1 | 2 | 2 |
|   Repetition Time, ms | 2000 | 1510 | 1550 | 2000 | 2000 | 2000 |
|   Echo Time, ms | 25 | 21 | 21 | 30 | 25 | 30 |
|   Flip Angle, ° | 80 | 90 | 90 | 75 | 80 | 75 |
|   Field of View, mm$^2$ | 220x220 | 205x205 | 205x205 | 240x240 | 220x220 | 240x240 |
|   Slice Number | 43 | 36 | 36 | 35 | 43 | 35 |
|   Slice-thickness, mm | 3.5 | 4.2 | 3.5 | 4.0 | 3.5 | 4.0 |
|   Gap, mm | 0 | 0 | 0 | 0 | 0 | 0 |
|   In-plane resolution, mm$^2$ | 3.4x3.4 | 3.2x3.2 | 3.2x3.2 | 3.0x3.0 | 3.4x3.4 | 3.0x3.0 |
|   Scan Duration, min | 5 | 9.3 | 9.4 | 7 | 5 | 7 |
|   Other Relevant Sequences | T1, Hypercapnic BOLD | T1, Hypercapnic BOLD, TOF-MRA | T1, Hypercapnic BOLD, T2-FLAIR | T1, T2-FLAIR, DWI | T1, Hypercapnic BOLD | T1 |

**Table 1 | Demographics and MRI sequence parameters of datasets used in this work.**

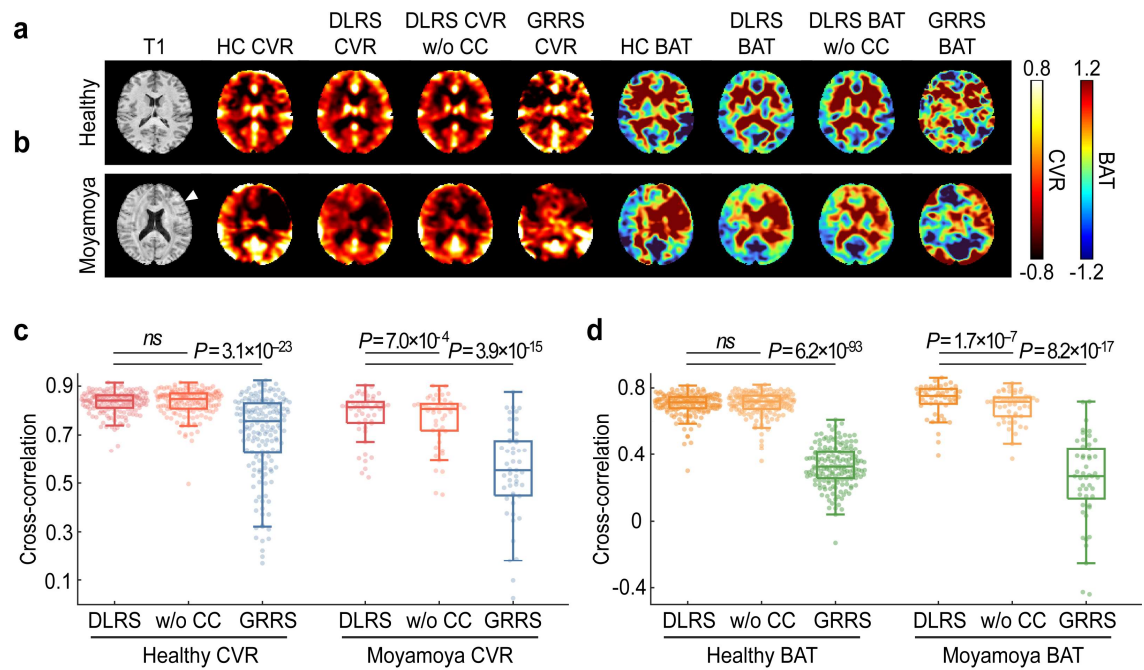

**Extended Data Fig. 1 | Examples and quantitative metrics of the ablation study. a, b,** Representative cerebrovascular reactivity (CVR) and bolus arrival time (BAT) images of a healthy volunteer **(a)**, and a Moyamoya disease patient **(b).** From left to right, the images are T1-weighted images, hypercapnic (HC) CVR, deep-learning resting-state (DLRS) CVR, ablated DLRS CVR, (global-regression resting-state) GRRS CVR, HC BAT, DLRS BAT, ablated DLRS BAT, and GRRS BAT. **c,** Pearson cross-correlation between RS CVR images and HC CVR images in healthy and Moyamoya disease participants. The ablated parametric maps revealed a lower correlation with the reference HC maps. **d,** Pearson cross-correlation between RS BAT images and HC BAT images.

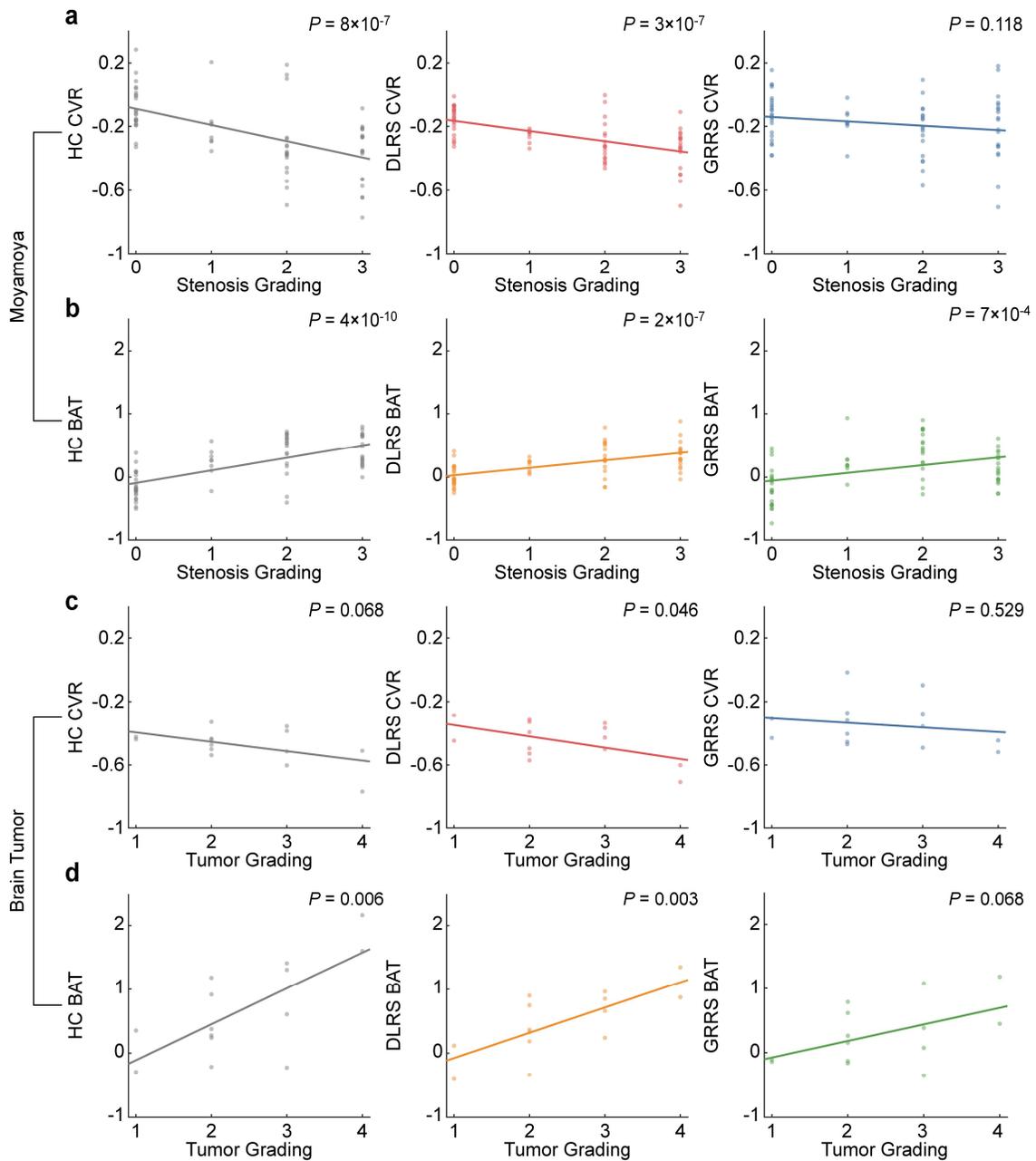

**Extended Data Fig. 2 | Associations between cerebrovascular physiological maps and clinical variables. a,** Scatter plots between regional CVR and arterial stenosis grade in Moyamoya patients, as rated in the middle cerebral artery (MCA). Plots are shown for hypercapnic cerebrovascular reactivity (HC CVR, left), deep-learning resting-state (DLRS CVR, middle) and global-regression resting-state (GRRS CVR, right). More severe stenosis was associated with lower CVR. **b,** Scatter plot between BAT and arterial stenosis grade. More severe stenosis was associated with longer BAT. **c,** Scatter plots between regional CVR and WHO tumor grade. **d,** Scatter plots between regional BAT and WHO tumor grade.

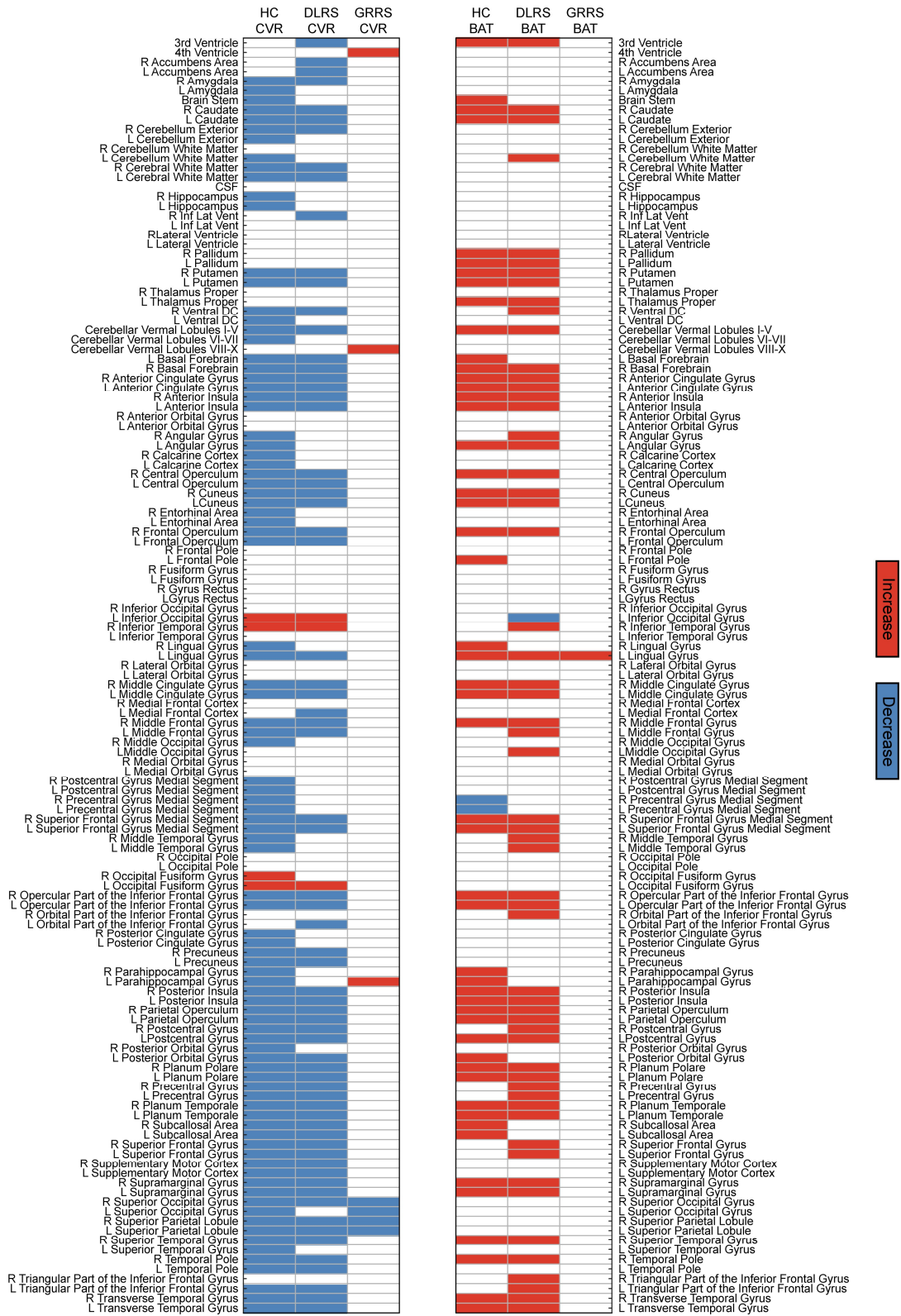

**Extended Data Fig. 3 | Age-related differences in deep-learning resting-state cerebrovascular reactivity (DLRS CVR) and bolus arrival time (DLRS BAT) across the lifespan.** From left to right, the heat maps represent the age differences associated with

hypercapnic (HC) CVR, DLRS CVR, global-regression resting-state (GRRS) CVR, HC BAT, DLRS BAT and GRRS BAT. Cell color denotes model type: linear increasing (Red) and linear decreasing (Blue). All models were added gender as covariate and controlled the false discovery rate (FDR, q<0.05).

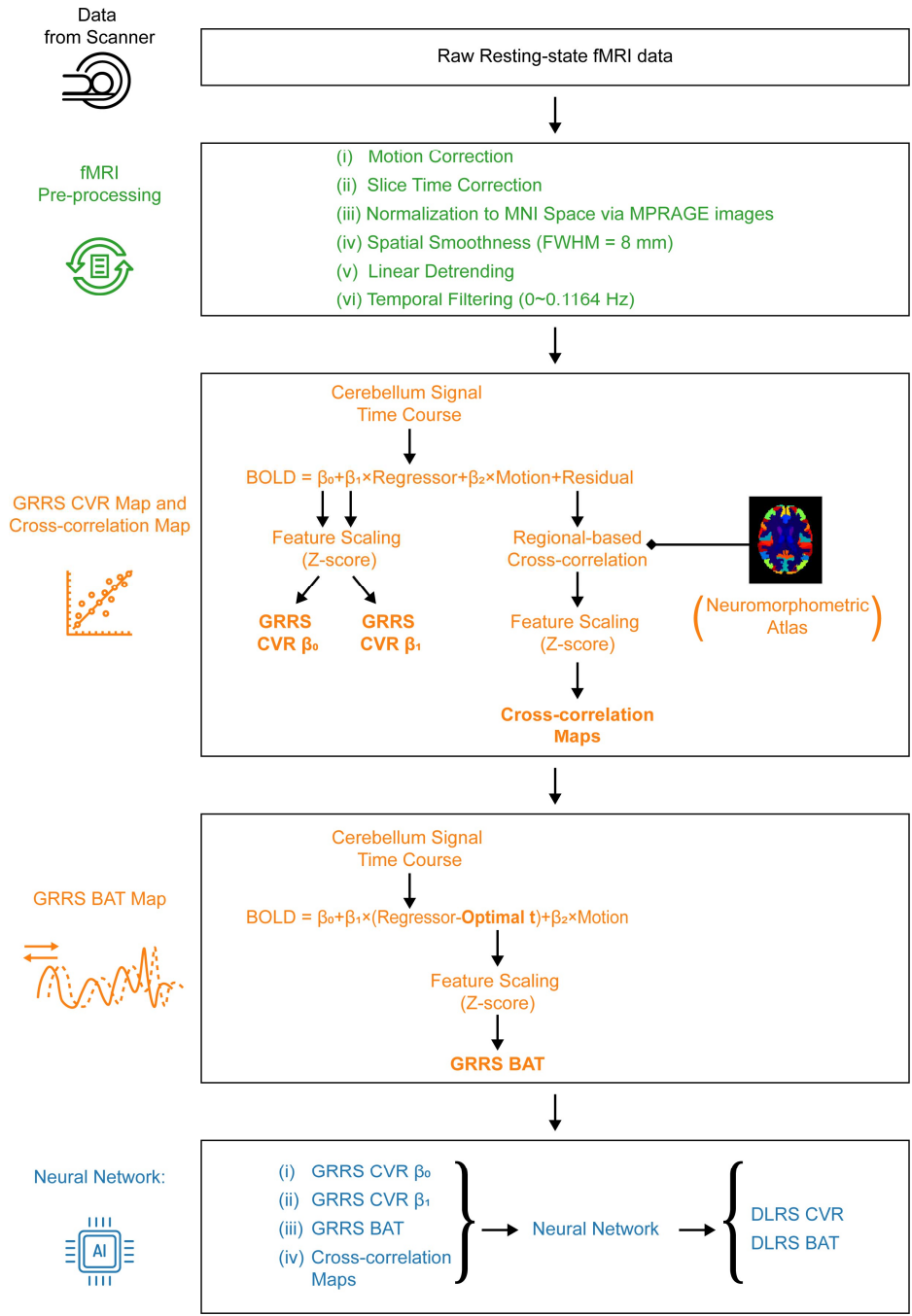

**Extended Data Fig. 4 | Illustration of the analysis pipelines of deep-learning resting-state cerebrovascular reactivity (DLRS CVR) and bolus arrival time (DLRS BAT) using resting-state fMRI data.**

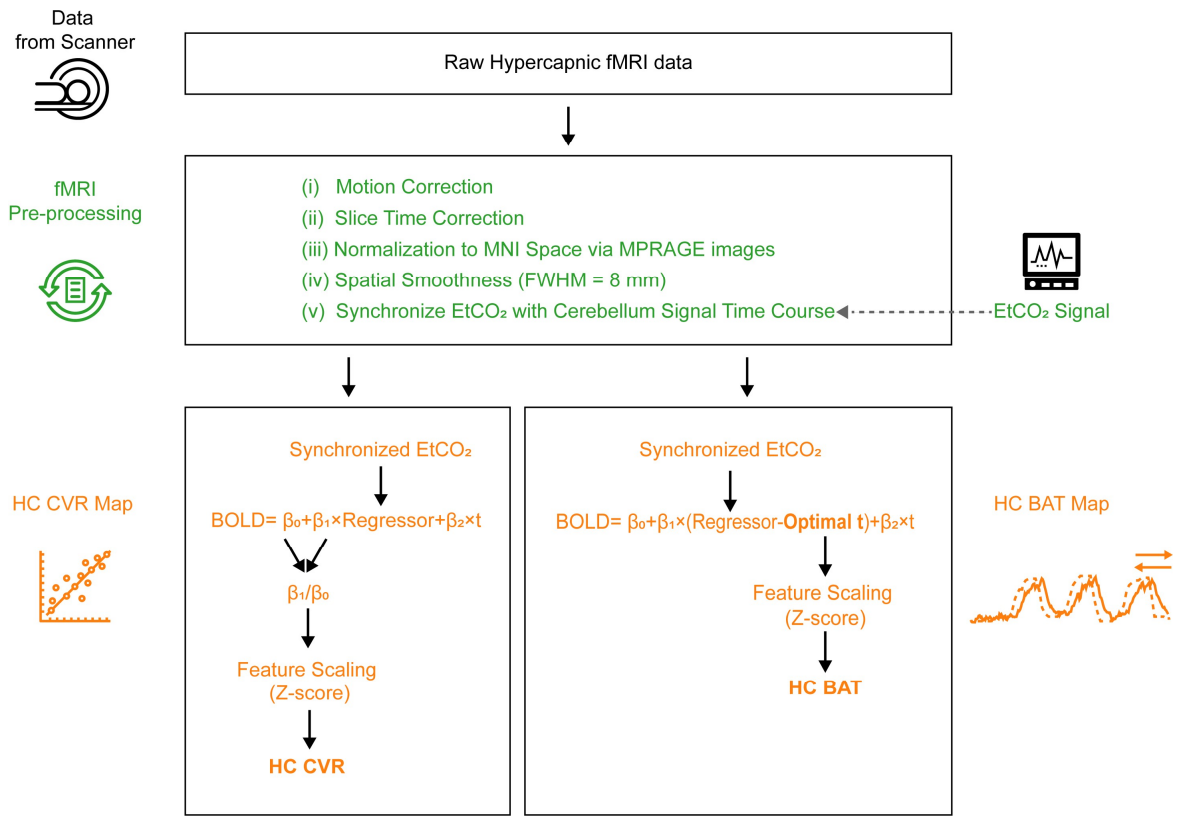

**Extended Data Fig. 5 | Illustration of the analysis pipelines of hypercapnic cerebrovascular reactivity (HC CVR) and bolus arrival time (HC BAT).**

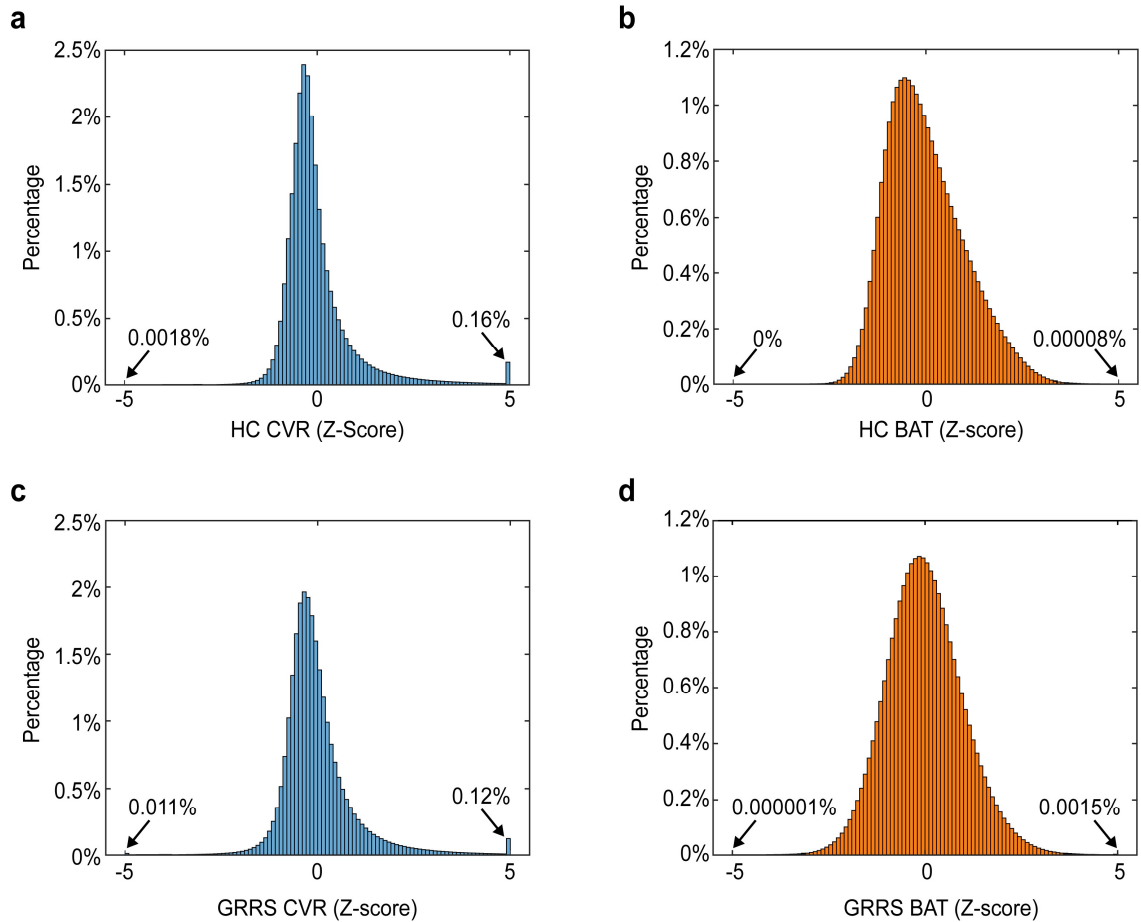

**Extended Data Fig. 6 | Histograms of cerebrovascular reactivity (CVR) and bolus arrival time (BAT) after imaging clipping to a range of less than 5. a,** Clipped distribution of hypercapnic (HC) CVR voxel-wise values after conversion to z-score. **b,** Clipped distribution of HC BAT voxel-wise values after conversion to z-score. **c,** Clipped distribution of global-regression resting-state (GRRS) CVR. **d,** Clipped distribution of GRRS BAT.

**Extended Data Table 1 | Architectural details of the deep-learning network used in this work.**

| Module | Block | Layer | Input Layer | Input Size | Num of Filters | Kernal Size | Stride | Padding | Module | Block | Layer | Input Layer | Input Size | Num of Filters | Kernal Size | Stride | Padding |
|---|---|---|---|---|---|---|---|---|---|---|---|---|---|---|---|---|---|
| Primary Encoder | P_Conv1 | Conv+ReLU+BN<br>Conv+ReLU+BN<br>MaxPool | Images<br>Prev<br>Prev | 3x96x112<br>64x96x112<br>64x96x112 | 64<br>64<br>1 | 3<br>3<br>2 | 1<br>1<br>2 | 1<br>1<br>0 | CVR Decoder | C_DeConv1 | Transpose Conv<br>Concatenate<br>Conv+ReLU+BN<br>Conv+ReLU+BN | P_Conv5+S_Conv5<br>P_Conv4+S_Conv4+Prev<br>Prev<br>Prev | 2048x6x7<br>1536x12x14<br>1536x12x14<br>512x12x14 | 512<br>-<br>512<br>512 | 2<br>-<br>3<br>3 | 2<br>-<br>1<br>1 | 0<br>-<br>1<br>1 |
| | P_Conv2 | Conv+ReLU+BN<br>Conv+ReLU+BN<br>MaxPool | P_Conv1<br>Prev<br>Prev | 64x48x56<br>128x48x56<br>128x48x56 | 128<br>128<br>1 | 3<br>3<br>2 | 1<br>1<br>2 | 1<br>1<br>0 | | C_DeConv2 | Transpose Conv<br>Concatenate<br>Conv+ReLU+BN<br>Conv+ReLU+BN | C_DeConv1<br>P_Conv3+S_Conv3+Prev<br>Prev<br>Prev | 512x12x14<br>768x24x28<br>768x24x28<br>256x24x28 | 256<br>-<br>256<br>256 | 2<br>-<br>3<br>3 | 2<br>-<br>1<br>1 | 0<br>-<br>1<br>1 |
| | P_Conv3 | Conv+ReLU+BN<br>Conv+ReLU+BN<br>MaxPool | P_Conv2<br>Prev<br>Prev | 128x24x28<br>256x24x28<br>256x24x28 | 256<br>256<br>1 | 3<br>3<br>2 | 1<br>1<br>2 | 1<br>1<br>0 | | C_DeConv3 | Transpose Conv<br>Concatenate<br>Conv+ReLU+BN<br>Conv+ReLU+BN | C_DeConv2<br>P_Conv2+S_Conv2+Prev<br>Prev<br>Prev | 256x24x28<br>384x48x56<br>384x48x56<br>128x48x56 | 128<br>-<br>128<br>128 | 2<br>-<br>3<br>3 | 2<br>-<br>1<br>1 | 0<br>-<br>1<br>1 |
| | P_Conv4 | Conv+ReLU+BN<br>Conv+ReLU+BN<br>MaxPool | P_Conv3<br>Prev<br>Prev | 256x12x14<br>512x12x14<br>512x12x14 | 512<br>512<br>1 | 3<br>3<br>2 | 1<br>1<br>2 | 1<br>1<br>0 | | C_DeConv4 | Transpose Conv<br>Concatenate<br>Conv+ReLU+BN<br>Conv+ReLU+BN | C_DeConv3<br>P_Conv1+S_Conv1+Prev<br>Prev<br>Prev | 128x48x56<br>192x96x112<br>192x96x112<br>64x96x112 | 64<br>-<br>64<br>64 | 2<br>-<br>3<br>3 | 2<br>-<br>1<br>1 | 0<br>-<br>1<br>1 |
| | P_Conv5 | Conv+ReLU+BN<br>Conv+ReLU+BN | P_Conv4<br>Prev | 512x6x7<br>1024x6x7 | 1024<br>1024 | 3<br>3 | 1<br>1 | 1<br>1 | | C_Output | Conv+Tanh | C_DeConv4 | 64x96x112 | 1 | 1 | 1 | 0 |
| Supplementary Encoder | S_Conv1 | Conv+ReLU+BN<br>Conv+ReLU+BN<br>MaxPool | Images<br>Prev<br>Prev | 133x96x112<br>64x96x112<br>64x96x112 | 64<br>64<br>1 | 3<br>3<br>2 | 1<br>1<br>2 | 1<br>1<br>0 | BAT Decoder | B_DeConv1 | Transpose Conv<br>Concatenate<br>Conv+ReLU+BN<br>Conv+ReLU+BN | P_Conv5+S_Conv5<br>P_Conv4+S_Conv4+Prev<br>Prev<br>Prev | 2048x6x7<br>1536x12x14<br>1536x12x14<br>512x12x14 | 512<br>-<br>512<br>512 | 2<br>-<br>3<br>3 | 2<br>-<br>1<br>1 | 0<br>-<br>1<br>1 |
| | S_Conv2 | Conv+ReLU+BN<br>Conv+ReLU+BN<br>MaxPool | S_Conv1<br>Prev<br>Prev | 64x48x56<br>128x48x56<br>128x48x56 | 128<br>128<br>1 | 3<br>3<br>2 | 1<br>1<br>2 | 1<br>1<br>0 | | B_DeConv2 | Transpose Conv<br>+Concatenate<br>Conv+ReLU+BN<br>Conv+ReLU+BN | B_DeConv1<br>P_Conv3+S_Conv3+Prev<br>Prev<br>Prev | 512x12x14<br>768x24x28<br>768x24x28<br>256x24x28 | 256<br>-<br>256<br>256 | 2<br>-<br>3<br>3 | 2<br>-<br>1<br>1 | 0<br>-<br>1<br>1 |
| | S_Conv3 | Conv+ReLU+BN<br>Conv+ReLU+BN<br>MaxPool | S_Conv2<br>Prev<br>Prev | 128x24x28<br>256x24x28<br>256x24x28 | 256<br>256<br>1 | 3<br>3<br>2 | 1<br>1<br>2 | 1<br>1<br>0 | | B_DeConv3 | Transpose Conv<br>Concatenate<br>Conv+ReLU+BN<br>Conv+ReLU+BN | B_DeConv2<br>P_Conv2+S_Conv2+Prev<br>Prev<br>Prev | 256x24x28<br>384x48x56<br>384x48x56<br>128x48x56 | 128<br>-<br>128<br>128 | 2<br>-<br>3<br>3 | 2<br>-<br>1<br>1 | 0<br>-<br>1<br>1 |
| | S_Conv4 | Conv+ReLU+BN<br>Conv+ReLU+BN<br>MaxPool | S_Conv3<br>Prev<br>Prev | 256x12x14<br>512x12x14<br>512x12x14 | 512<br>512<br>1 | 3<br>3<br>2 | 1<br>1<br>2 | 1<br>1<br>0 | | B_DeConv4 | Transpose Conv<br>Concatenate<br>Conv+ReLU+BN<br>Conv+ReLU+BN | B_DeConv3<br>P_Conv1+S_Conv1+Prev<br>Prev<br>Prev | 128x48x56<br>192x96x112<br>192x96x112<br>64x96x112 | 64<br>-<br>64<br>64 | 2<br>-<br>3<br>3 | 2<br>-<br>1<br>1 | 0<br>-<br>1<br>1 |
| | S_Conv5 | Conv+ReLU+BN<br>Conv+ReLU+BN | S_Conv4<br>Prev | 512x6x7<br>1024x6x7 | 1024<br>1024 | 3<br>3 | 1<br>1 | 1<br>1 | | B_Output | Conv+Tanh | B_DeConv4 | 64x96x112 | 1 | 1 | 1 | 0 |

Conv: Convolution, ReLU: Rectified linear unit, BN: Batch normalization, Tanh: Hyperbolic tangent, Prev: Previous layer output.

**Extended Data Table 1 | Architectural details of the deep-learning network used in this work.**

## Acknowledgments

This work was supported by grants from the National Institute of Health (R01 AG064792, RF1 AG071515, R01 NS106711, R01 NS106702, UF1NS100588, RF1 AG006265, R01 DC005375, R01 DC015466, P41 EB031771, and S10 OD021648).

## Author Contributions

X.H. and H.L. conceived and designed the study. X.H., P.G., P.W. and V.M.P. designed the deep-learning network. X.H., P.L., D.D.L, H.F., Y.L., Z.W., Z.L., D.J., J.J., C.K., J.J.P., J.H., M.C.P., B.P.T., and B.G.W. contributed to the acquisition and analysis of data. D.C.P., V.M.P., A.E.H., and H.L. coordinated and oversaw the study. All authors participated in discussion and interpretation of the data. X.H. and H.L. wrote the manuscript with input from all authors.

## Ethics Declarations

All authors declare that they have no conflicts of interest.